\documentclass[aps,twocolumn,superscriptaddress,showkeys,showpacs]{revtex4}
\usepackage{amssymb}
\usepackage{amsfonts}
\usepackage{amsmath}
\usepackage{amsthm}
\usepackage{epsfig}
\usepackage{color} 
\usepackage{subfigure}
\usepackage{bbm}

\begin{document}

\title{Coherence-resonance chimeras in a network of excitable elements}

\author{Nadezhda Semenova}
\affiliation{Department of Physics, Saratov State University, Astrakhanskaya str.~83, 410012 Saratov, Russia}
\author{Anna Zakharova} 
\email[corresponding author: ]{anna.zakharova@tu-berlin.de}
\affiliation{Institut f{\"u}r Theoretische Physik, Technische Universit\"at Berlin, Hardenbergstra\ss{}e 36, 10623 Berlin, Germany}
\author{Vadim Anishchenko}
\affiliation{Department of Physics, Saratov State University, Astrakhanskaya str.~83, 410012 Saratov, Russia}
\author{Eckehard Sch{\"o}ll}
\affiliation{Institut f{\"u}r Theoretische Physik, Technische Universit\"at Berlin, Hardenbergstra\ss{}e 36, 10623 Berlin, Germany}

\date{\today}

\begin{abstract}
We demonstrate that chimera behavior can be observed in nonlocally coupled networks of excitable systems in the presence of noise. This phenomenon is distinct from classical chimeras, which occur in deterministic oscillatory systems, and it combines temporal features of coherence resonance, i.e., the constructive role of noise, 
and spatial properties of chimera states, i.e., coexistence of spatially coherent and incoherent domains in a network of identical elements. Coherence-resonance chimeras are associated with alternating switching of the location of coherent and incoherent domains, which might be relevant in neuronal networks.
\end{abstract}



\maketitle 

Chimera states are intriguing spatio-temporal patterns made up of spatially separated domains of synchronized (spatially coherent) and desynchronized (spatially incoherent) behavior, arising in networks of identical units. 
Originally discovered in a network of phase oscillators with a simple symmetric non-local coupling scheme \cite{KUR02a,ABR04}, this sparked a tremendous activity
of theoretical investigations \cite{PAN15,ABR08,SET08,LAI09,MOT10,MAR10,OLM10,BOR10,SHE10,SEN10a,WOL11,LAI11,OME11,OME12,OME13,NKO13,HIZ13,SET13,SET14,YEL14,BOE15,BUS15,OME15,OME15a, ASH14}. The first experimental evidence on chimera states was presented only one decade after their theoretical discovery \cite{HAG12,TIN12,MAR13,LAR13,KAP14,WIC13,WIC14,SCH14a,GAM14,ROS14a,LAR15}. In real-world systems chimera states might play a role, e.g., in power grids~\cite{MOT13a}, in social systems~\cite{GON14}, in the unihemispheric sleep of birds and dolphins~\cite{RAT00}, or in epileptic seizures~\cite{ROT14}. In the context of the latter two applications it is especially relevant to explore chimera states in neuronal networks under conditions of excitability. 
However, while chimera states have previously been reported for neuronal networks in the oscillatory regime, e.g., in the FitzHugh-Nagumo system~\cite{OME13}, they have not been detected 
in the excitable regime even for specially prepared initial conditions \cite{OME13}. Therefore, the existence of chimera states for excitable elements remains unresolved. 

One of the challenging issues concerning chimera states is their behavior in the presence of random fluctuations, which are unavoidable in real-world systems.
The robustness of chimeras with respect to external noise has been studied only very recently \cite{LOO15}. An even more intriguing question is whether 
the constructive role of noise in nonlinear systems, manifested for example in the counter-intuitive increase of temporal coherence due to noise in \textit{coherence resonance} \cite{HU93a,PIK97,NEI97,LIN04}, 
can be combined with the chimera behavior in spatially extended systems and networks.
Coherence resonance, originally discovered for excitable systems like the FitzHugh-Nagumo model, implies that noise-induced oscillations become more regular for an optimum intermediate value of noise intensity. A question naturally arising in this context is whether noise can also have a beneficial effect on chimera states. No evidence for the constructive role of noise for chimeras has been previously provided. Therefore, an important issue we aim to address here is to establish a connection between two intriguing counter-intuitive phenomena which have been previously studied independently in different scientific communities: coherence resonance and chimera states. 

In this Letter we investigate an effect which combines coherence resonance and chimera states in a network of nonlocally coupled excitable elements. We demonstrate that chimera behavior can be observed in excitable systems and not only in oscillatory systems and show that the presence of noise is a crucial condition for this case. Moreover, we uncover the constructive role of noise for chimera states and detect a novel type of coherence resonance, which we call {\em coherence-resonance chimeras}. In these spatio-temporal patterns coherence resonance is associated with spatially coherent and incoherent behavior, rather than purely temporal coherence or regularity measured by the correlation time. Since we consider a paradigmatic model for neural excitability in a realistic noisy environment, we expect wide-range applications of our results to neuronal networks in general. Moreover, the noise-based control mechanism we propose here reveals an alternative direction for chimera control complementary to recent deterministic control schemes~\cite{SIE14c,BIC15}.


The FitzHugh-Nagumo (FHN) oscillator is a paradigmatic model for excitable systems, originally suggested to characterize the spiking behaviour of neurons \cite{FIT61,NAG62,SCO75,KLI15}. Its fields of application range from neuroscience and biological processes \cite{LIN04,CIS03} to optoelectronic \cite{ROS11a} and chemical \cite{SHI04} oscillators and nonlinear electronic circuits \cite{HEI10}. 
We consider a ring of $N$ nonlocally coupled FHN oscillators in the presence of Gaussian white noise:
\begin{equation}\label{eq:ring_fhn}
\begin{array}{c}
\varepsilon\frac{du_i}{dt}=u_i-\frac{u^3_i}{3}-v_i +\frac{\sigma}{2R}\sum\limits_{j=i-R}^{i+R} [b_{uu}(u_j-u_i)\\
+b_{uv}(v_j-v_i)], \\
\frac{dv_i}{dt}=u_i+a_i+ \frac{\sigma}{2R}\sum\limits_{j=i-R}^{i+R} [b_{vu}(u_j-u_i) \\
+b_{vv}(v_j-v_i)] + \sqrt{2D} \xi_{i}(t),
\end{array}
\end{equation}
where $u_i$ and $v_i$ are the activator and inhibitor variables, respectively, $i=1,...,N$ and all indices are modulo $N$, $\varepsilon>0$ is a small parameter responsible for the time scale separation of fast activator and slow inhibitor, $a_i$ defines the excitability threshold. For an individual FHN element it determines whether the system is excitable ($|a_i|>1$), or oscillatory ($|a_i|<1$). In the following we assume that all elements are in the excitable regime close to the threshold ($a_{i}\equiv a=1.001$), $\sigma$ is the coupling strength, $R$ is the number of nearest neighbours and $r=R/N$ is the coupling range. The form of the coupling of Eq.~(\ref{eq:ring_fhn}) is inspired from neuroscience \cite{OME13,KOZ98,HUL04,HEN11}, where strong interconnections between neurons are found within a range $R$, but much fewer connections exist at longer distances. Further, $\xi_i(t) \in \mathbb{R}$ is Gaussian white noise, i.e., $\langle \xi_i (t) \rangle \!=\! 0$ and $\langle \xi_i (t) \xi_j(t') \rangle \!=\!  \delta_{ij} \delta(t-t'), ~\forall i,j$, 
and $D$ is the noise intensity. 
Eq.~(\ref{eq:ring_fhn}) contains not only direct, but also cross couplings between activator ($u$) and inhibitor ($v$) variables, which is modelled by a rotational coupling matrix \cite{OME13}
$b_{uu}=b_{vv}=\cos\phi$, $b_{uv}= -b_{vu}= \sin\phi$ with $\phi\in[-\pi;\pi)$. In this Letter we fix the parameter $\phi=\pi/2-0.1$ for which chimeras have been found in the deterministic oscillatory regime~\cite{OME13}. 
In the excitable regime ($|a|>1$) a single FHN oscillator rests in a locally stable steady state and upon excitation by noise beyond a threshold emits a spike. With increasing noise the excitation across threshold occurs more frequently, and thus the interspike intervals become more regular. On the other hand, with increasing noise the deterministic spiking dynamics becomes smeared out. The best temporal regularity is observed for an optimum intermediate value of noise intensity and the corresponding counter-intuitive phenomenon is known as coherence resonance \cite{HU93a,PIK97,NEI97}. Such behavior has been shown theoretically and experimentally in a variety of excitable systems, like lasers with saturable absorber \cite{DUB99}, optical feedback \cite{GIA00,AVI04}, and optical injection \cite{ZIE13}, semiconductor superlattices \cite{HIZ06,HUA14}, or neural systems \cite{PIK97,LIN04,JAN03} and recently in non-excitable systems as well \cite{USH05,ZAK10a,ZAK13,GEF14,SEM15}. Different temporal correlation measures can be used to detect coherence resonance \cite{PIK97,ROS09b}, see supplemental material.  

On the other hand, the spatial coherence and incoherence of chimera states can be characterized by a local order parameter \cite{OME11,WOL11a}:
\begin{equation}
Z_k=\Big|\frac{1}{2\delta_Z}\sum\limits_{|j-k|\leq\delta_Z} e^{i \Theta_j}\Big|, \ \ \ k=1,\dots N
\end{equation}
where the geometric phase of the $j$th oscillator is defined by $\Theta_j=arctan(v_j/u_j)$ \cite{OME13} and $Z_k \approx 1$ and $Z_k<1$ indicate coherence and incoherence, respectively.

\begin{figure}[htbp]
\includegraphics[width=0.8\linewidth]{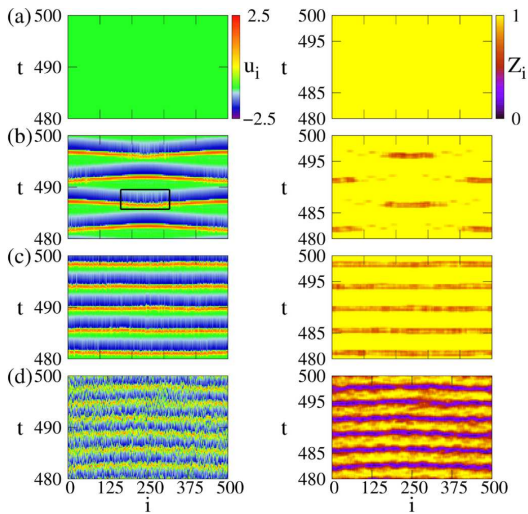}
\caption[]{Space-time plots of activator $u_i$ (left column) and local order parameter $Z_i$ (right column, $\delta_Z=25$) for different noise intensities (a) $D=0$: steady state, (b) $D=0.0002$: coherence-resonance chimera (initial conditions: randomly distributed on the circle $u^2 + v^2 = 4$), (c) $D=0.0004$: incoherent in space but periodic in time, (d) $D=0.1$: incoherent in space and time. Parameters: $\varepsilon=0.05$, $a=1.001$, $\sigma=0.4$, $r=0.12$.}
\label{fig:main_types}
\end{figure}

Depending on the noise intensity $D$, we find four distinct regimes which are shown as space-time plots color-coded by the variable $u_{i}$ and by the local order parameter $Z_{i}$ in Fig.~\ref{fig:main_types}. If the noise intensity is low ($D < 0.000062$), the network rests in a homogeneous steady state [Fig.~\ref{fig:main_types}(a)]. For intermediate values of noise intensity $0.000062 \le D \le 0.000325$ we find novel spatio-temporal spiking patterns, which combine features of chimera states and coherence resonance; therefore, we call them \textit{coherence-resonance chimeras} [Fig.~\ref{fig:main_types}(b)]. 
These patterns are characterized by the coexistence of two different domains separated in space, where one part of the network is spiking coherently in space while the other exhibits incoherent spiking, i.e., the spiking of neighbouring nodes is uncorrelated. From the incoherent domain in the center of the space-time plot (marked with a black rectangle) two very fast counterpropagating excitation waves emanate, and as they propagate they become coherent and as they meet again on the antipodal position on the ring they annihilate. Subsequently, at that position around $i=500$, another incoherent domain is born, which again generates two fast counterpropagating coherent excitation waves, and so on. In order to quantify coherence and incoherence for this pattern we calculate the local order parameter $Z_{i}$ (right panel in Fig.~\ref{fig:main_types}(b)). It can be clearly seen that the islands of desynchronization corresponding to the incoherent domains are characterized by values of the order parameter noticeably below unity (dark patches). Surprisingly, in this stochastic chimera regime, the width of the incoherent domain remains fixed, a typical feature of classical chimera states found in phase oscillator networks. However, as it often occurs in stochastic systems, noise induces switching behavior in time. Here, driven by the interplay of Gaussian white noise and nonlocality of the coupling, the incoherent domain of the chimera pattern switches periodically its position on the ring. Such an interchange between the coherent and the incoherent domains of the chimera state is crucial for the understanding of unihemispheric sleep, where the synchronization of neurons is known to switch between hemispheres of the brain \cite{HAU15}. Previously, such alternating chimera behavior has been reported for a deterministic oscillatory medium with nonlinear global coupling 
\cite{HAU15}. Here, we show that the alternating behavior can be caused in excitable media by stochasticity, which is inherent to real-world systems. Therefore, we propose that coherence-resonance chimeras which we uncover for a network of neuronal oscillators in stochastic environment, may provide a bridge 
between chimera states and unihemispheric sleep.

Interestingly, coherence-resonance chimeras appear to be a persistent phenomenon, which continues to exist for at least $T_{int}=10^5$ dimensionless integration time units, which corresponds to $\approx 35000$ intrinsic periods. This discloses the constructive role of noise for the considered pattern in contrast to amplitude chimeras, which tend to have shorter lifetimes monotonically decreasing with increasing noise \cite{LOO15}.

Strong noise destroys coherence-resonance chimeras. For noise intensity $D>0.000325$ the system Eq.~(\ref{eq:ring_fhn}) is completely incoherent in space but still very regular (approximately periodic) in time [Fig.~\ref{fig:main_types}(c)]. In the case of even stronger noise, for instance $D=0.1$ [Fig.~\ref{fig:main_types}(d)], the behavior becomes incoherent both in space and in time.
Therefore, coherence-resonance chimeras appear for intermediate values of noise intensity, which is a characteristic signature of coherence resonance. Note that coherence resonance chimeras occur in a network at much lower values of the noise intensity than coherence resonance in a single FHN system; this is due to the strong coupling of each element with its neighbors.
\begin{figure}[htbp]
\includegraphics[width=0.6\linewidth]{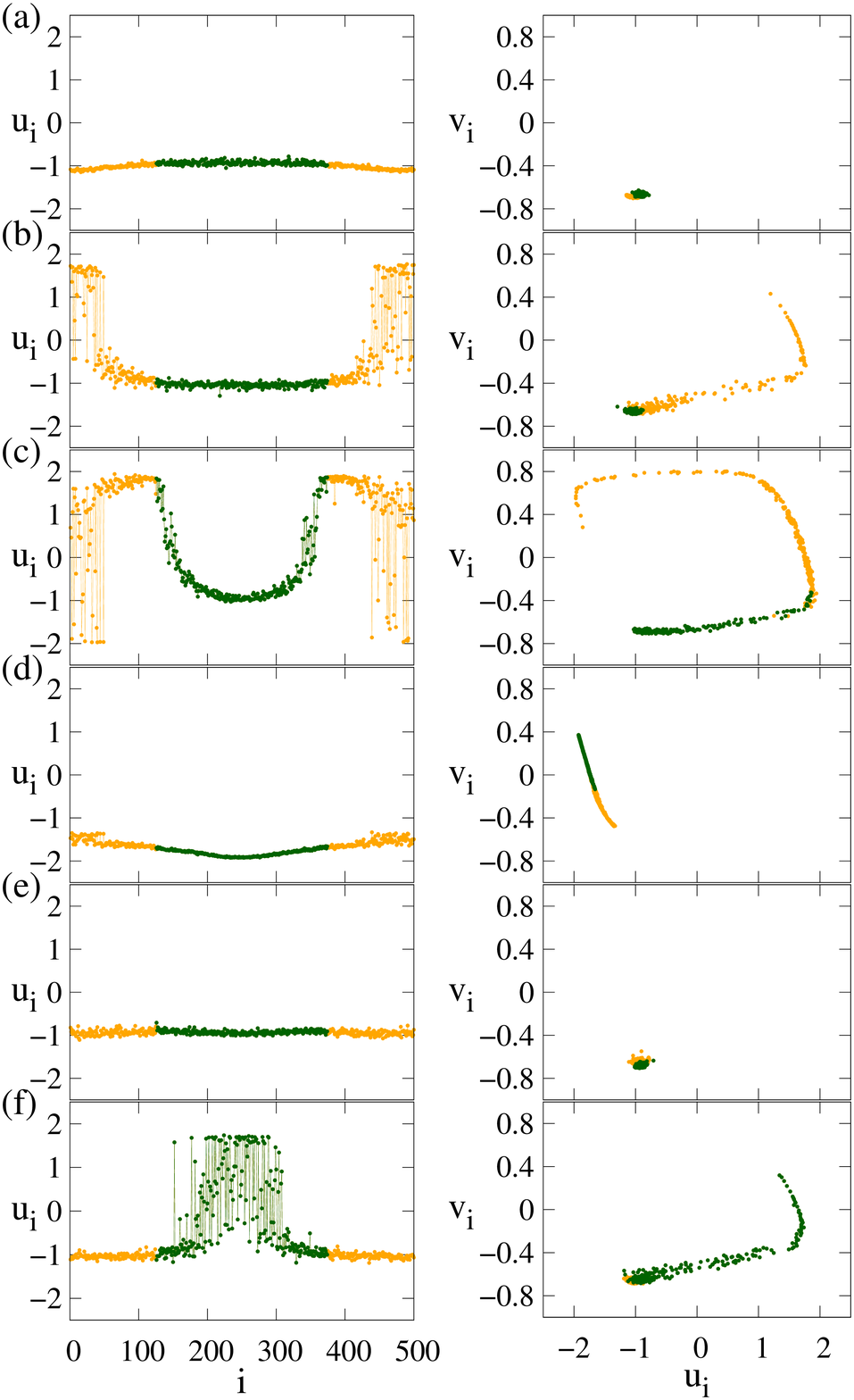}
\caption[~~~]{Time evolution of coherence-resonance chimera: Snapshots (left column) and corresponding phase space (right column) for
(a) $t=976.89$, (b) $t=977.75$, (c) $t=978.13$, (d) $t=979.81$, (e) $t=981.87$, (f) $t=982.48$.
 Other parameters: $\varepsilon=0.05$, $a=1.001$, $D=0.0002$, $\sigma=0.4$, $r=0.2$.}
\label{fig:time_evolution}
\end{figure}

To provide more insight into the coherence-resonance chimera state in Fig.~\ref{fig:main_types}(b), we investigate a temporal sequence of snapshots of the variable $u_{i}$ and phase portraits in the ($u_{i}, v_{i}$)-plane (Fig.~\ref{fig:time_evolution}). The middle nodes from $i=125$ to $i=375$ are marked in green (dark). We start with the state where all the elements of the network are located close to the steady state $u_{i}\approx-1$ (panel a). A little bit later (panel b) one group of elements (on the left and the right side in the snapshot) incoherently starts the excursion in phase space (phase portrait in panel (b)), and as the excitation rapidly moves to the middle part from both sides, it becomes more and more coherent (panel c). This phase in the time evolution corresponds to spiking. 
Note that the nodes from the incoherent domain start their journey in the phase space first (desynchronized spiking) while the nodes from the coherent domain catch up later but more synchronously. Next, all the FHN oscillators jump back to the left branch of the activator nullcline in phase space and return along the nullcline slowly and rather synchronously to the steady state [Fig.~\ref{fig:time_evolution}(d)]. The incoherent bunch of nodes returns to the steady state first, but is scattered around the steady state for a while (see right panel e); at the same time the coherent part is catching up, but coherently. Therefore, the coherent nodes are more likely to cross the threshold and spike again - due to noise they do it in an incoherent way.  Further, the steps described above repeat, however, with the coherent and the incoherent domains interchanged. This explains why the coherent and incoherent spiking alternates between the two groups, shown in orange and in green, respectively.

\begin{figure}[htbp]
\includegraphics[width=0.6\linewidth]{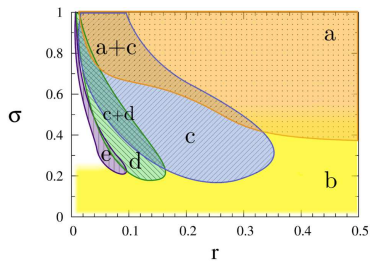}
\caption[~~~]{Dynamic regimes in the $(r,\sigma)$ parameter plane: (a) steady state (orange dotted),
(b) incoherent in space and periodic in time (yellow plain), 
(c) coherence-resonance (CR) chimera with one incoherent domain (blue cross-hatched) 
(d) CR chimera with two incoherent domains (green cross-hatched) 
(e) CR chimera with three incoherent domains (purple vertically hatched). 
Multistability is also indicated. Other parameters: $\varepsilon=0.05$, $a=1.001$, $D=0.0002$, $N=500$.}
\label{fig:param_plane}
\end{figure}

The number of the incoherent domains can be increased by decreasing the coupling range $r$ for fixed value of the coupling strength 
$\sigma$, a typical feature of ``classical chimera states'', cf. e.g. \cite{OME11,HAG12,OME13,ZAK14,OME15a}. For coherence-resonance chimeras with two and three incoherent domains see the supplemental material.
To gain an overview of the different regimes we fix the values of parameters $\varepsilon$, $a$, $D$, $N$,  
and tune $r$ and $\sigma$ (Fig.~\ref{fig:param_plane}). 
Strong coupling and a large number of nearest neighbors force the network to rest in the homogeneous steady state (region a). For weaker coupling and almost the whole range of $r$ values we detect spiking patterns which are approximately periodic in time and incoherent in space (region b). Coherence-resonance chimeras occur above a certain threshold $\sigma\approx0.2$. Depending on the coupling range $r$ we find coherence-resonance chimeras with one, two, and three incoherent domains (regions c, d and e, respectively). Note that near the borders of the different regimes multistability is observed (regions a+c and c+d), and the initial conditions determine the particular pattern. 

\begin{figure}[htbp]
\includegraphics[width=0.8\linewidth]{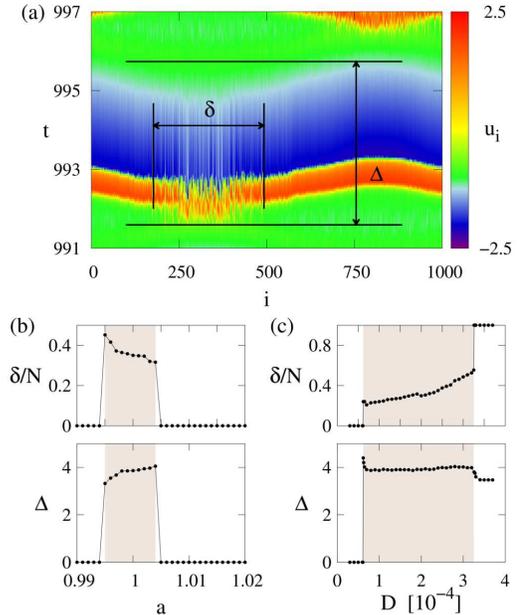}
\caption[~~~]{Characterization of CR chimera: (a) Space-time plot defining active time $\Delta$ and size $\delta$ of the incoherent domain. (b),(c) Dependence of $\delta/N$ and $\Delta$ upon excitation threshold $a$ for $D = 0.0002$ (a) and upon noise intensity $D$ for $a = 1.001$ (c). Gray region corresponds to the existence of CR chimeras.
Other parameters: $\varepsilon=0.05$, $N=1000$, $\sigma=0.4$, $r = 0.2$.}
\label{fig:impact_D}
\label{fig:impact_a}
\label{fig:notation}
\end{figure}

To deepen our understanding of coherence-resonance chimeras we analyze the impact of the excitation threshold $a$. Since chimera states in the deterministic FHN model have been previously observed only in the oscillatory regime for $|a|<1$~\cite{OME13}, we investigate if coherence-resonance chimeras are sensitive to the choice of $a$. 
For that purpose we consider two characteristic quantities: (i) the normalized size of the incoherent domain $\delta/N$, where $\delta$ is the number of elements in the incoherent domain [Fig.~\ref{fig:notation}(a)];
(ii) the active time span of the chimera $\Delta$, which measures the time from the excitation of the first node belonging to the incoherent domain till the return of the last node to the rest state [Fig.~\ref{fig:notation}~(a)].
This is analogous to the pulse duration for the single FHN model \cite{PIK97}, but takes into account that different nodes spike at distinct moments of time depending on the domain.
Our results show that for increasing $a$ the incoherent domain size $\delta/N$ shrinks (top panel in Fig.~\ref{fig:impact_a}(b)) and the active time span $\Delta$ increases (bottom panel). Interestingly, coherence-resonance chimeras occur for both oscillatory and excitable regimes of FHN oscillators, but they exist only for a restricted interval of the threshold parameter $a$ (shaded region $0.995 \le a \le 1.004$ in Fig.~\ref{fig:impact_a}(b)). To the left of this interval the dynamics is completely synchronized in space and periodic in time, while to the right the patterns are incoherent in space and periodic in time (similar to Fig.~\ref{fig:main_types}(c)). Fig.~\ref{fig:impact_a}(c) shows that $\delta/N$ increases with noise intensity $D$ 
(top panel of Fig.~\ref{fig:impact_a}(c)), while $\Delta$ is independent of $D$ within the interval of existence of the coherence-resonance chimeras $0.000062 \le D \le 0.000325$ (bottom panel). 

In conclusion, noise plays a crucial role for coherence-resonance chimera patterns for two main reasons: on the one hand it induces the pattern and on the other hand allows to control it. The coherence-resonance attribute of this pattern discloses the first aspect: coherence-resonance chimeras appear for intermediate values of noise intensity. However, this can also be viewed from the control perspective: by properly choosing the noise intensity we achieve the desired regime of the network: steady state, coherence-resonance chimera, or other patterns. 
Indeed, by fine tuning the noise intensity we can adjust the size $\delta$ of the incoherent domain of the chimera pattern. While the active time span remains fixed for all noise intensities within the interval of existence, the size of the incoherent domain $\Delta$ essentially grows with increasing noise intensity. An important aspect of our work is also that these novel coherence-resonance chimeras in a neural network under the influence of noise exhibit alternating chimera behavior, i.e., the coherent and incoherent domains switch position periodically, which might offer, for instance, a natural explanation of the phenomenon of unihemispheric sleep.

We acknowledge Tatyana Vadivasova, Iryna Omelchenko and Vladimir Semenov for valuable discussions.   
This work was supported by DFG in the framework of SFB 910 and by the Russian Foundation for Basic
Research (Grants No. 15-02-02288 and 14-52-12002). 

\bibliographystyle{prwithtitle}

\begin{thebibliography}{100}

\bibitem{KUR02a}
Y. Kuramoto and D. Battogtokh, Nonlin. Phen. in Complex Sys. {\bf 5},  380
  (2002).

\bibitem{ABR04}
D.~M. Abrams and S.~H. Strogatz, Phys.~Rev.~Lett. {\bf 93},  174102  (2004).

\bibitem{PAN15}
M.~J. Panaggio and D.~M. Abrams, Nonlinearity {\bf 28},  R67  (2015).

\bibitem{ABR08}
D.~M. Abrams, R.~E. Mirollo, S.~H. Strogatz, and D.~A. Wiley, Phys. Rev. Lett.
  {\bf 101},  084103  (2008).

\bibitem{SET08}
G.~C. Sethia, A. Sen, and F.~M. Atay, Phys.~Rev.~Lett. {\bf 100},  144102
  (2008).

\bibitem{LAI09}
C.~R. Laing, Physica D {\bf 238},  1569  (2009).

\bibitem{MOT10}
A.~E. Motter, Nature Phys. {\bf 6},  164  (2010).

\bibitem{MAR10}
E.~A. Martens, C.~R. Laing, and S.~H. Strogatz, Phys. Rev. Lett. {\bf 104},
  044101  (2010).

\bibitem{OLM10}
S. Olmi, A. Politi, and A. Torcini, Europhys. Lett. {\bf 92},  60007  (2010).

\bibitem{BOR10}
G. Bordyugov, A. Pikovsky, and M. Rosenblum, Phys. Rev. E {\bf 82},  035205
  (2010).

\bibitem{SHE10}
J.~H. Sheeba, V.~K. Chandrasekar, and M. Lakshmanan, Phys. Rev. E {\bf 81},
  046203  (2010).

\bibitem{SEN10a}
A. Sen, R. Dodla, G. Johnston, and G.~C. Sethia,  in {\em Complex Time-Delay
  Systems}, Vol.~16 of {\em Understanding Complex Systems}, edited by F.~M.
  Atay (Springer, Berlin, 2010), pp.\ 1--43.

\bibitem{WOL11}
M. Wolfrum and O.~E. Omel'chenko, Phys. Rev. E {\bf 84},  015201  (2011).

\bibitem{LAI11}
C.~R. Laing, Physica D {\bf 240},  1960  (2011).

\bibitem{OME11}
I. Omelchenko, Y. Maistrenko, P. H{\"o}vel, and E. Sch{\"o}ll, Phys. Rev. Lett.
  {\bf 106},  234102  (2011).

\bibitem{OME12}
I. Omelchenko, B. Riemenschneider, P. H{\"o}vel, Y. Maistrenko, and E.
  Sch{\"o}ll, Phys. Rev.~E {\bf 85},  026212  (2012).

\bibitem{OME13}
I. Omelchenko, O.~E. Omel'chenko, P. H{\"o}vel, and E. Sch{\"o}ll, Phys. Rev.
  Lett. {\bf 110},  224101  (2013).

\bibitem{NKO13}
S. Nkomo, M.~R. Tinsley, and K. Showalter, Phys. Rev. Lett. {\bf 110},  244102
  (2013).

\bibitem{HIZ13}
J. Hizanidis, V. Kanas, A. Bezerianos, and T. Bountis, Int. J. Bifurcation
  Chaos {\bf 24},  1450030  (2014).

\bibitem{SET13}
G.~C. Sethia, A. Sen, and G.~L. Johnston, Phys. Rev. E {\bf 88},  042917
  (2013).

\bibitem{SET14}
G.~C. Sethia and A. Sen, Phys. Rev. Lett. {\bf 112},  144101  (2014).

\bibitem{YEL14}
A. Yeldesbay, A. Pikovsky, and M. Rosenblum, Phys. Rev. Lett. {\bf 112},
  144103  (2014).

\bibitem{BOE15}
F. B{\"o}hm, A. Zakharova, E. Sch{\"o}ll, and K. L{\"u}dge, Phys. Rev. E {\bf
  91},  040901 (R)  (2015).

\bibitem{BUS15}
A. Buscarino, M. Frasca, L.~V. Gambuzza, and P. H{\"o}vel, Phys. Rev.~E {\bf
  91},  022817  (2015).

\bibitem{OME15}
I. Omelchenko, A. Provata, J. Hizanidis, E. Sch{\"o}ll, and P. H{\"o}vel, Phys.
  Rev. E {\bf 91},  022917  (2015).

\bibitem{OME15a}
I. Omelchenko, A. Zakharova, P. H{\"o}vel, J. Siebert, and E. Sch{\"o}ll, Chaos
  {\bf 25},  083104  (2015).

\bibitem{ASH14}
P. Ashwin and O. Burylko, Chaos {\bf 25},  013106  (2015).

\bibitem{HAG12}
A.~M. Hagerstrom, T.~E. Murphy, R. Roy, P. H{\"o}vel, I. Omelchenko, and E.
  Sch{\"o}ll, Nature Phys. {\bf 8},  658  (2012).

\bibitem{TIN12}
M.~R. Tinsley, S. Nkomo, and K. Showalter, Nature Phys. {\bf 8},  662  (2012).

\bibitem{MAR13}
E.~A. Martens, S. Thutupalli, A. Fourriere, and O. Hallatschek, Proc. Nat.
  Acad. Sciences {\bf 110},  10563  (2013).

\bibitem{LAR13}
L. Larger, B. Penkovsky, and Y. Maistrenko, Phys. Rev. Lett. {\bf 111},  054103
   (2013).

\bibitem{KAP14}
T. Kapitaniak, P. Kuzma, J. Wojewoda, K. Czolczynski, and Y. Maistrenko, Sci.
  Rep. {\bf 4},  6379  (2014).

\bibitem{WIC13}
M. Wickramasinghe and I.~Z. Kiss, PLoS ONE {\bf 8},  e80586  (2013).

\bibitem{WIC14}
M. Wickramasinghe and I.~Z. Kiss, Phys. Chem. Chem. Phys. {\bf 16},  18360
  (2014).

\bibitem{SCH14a}
L. Schmidt, K. Sch{\"o}nleber, K. Krischer, and V. Garcia-Morales, Chaos {\bf
  24},  013102  (2014).

\bibitem{GAM14}
L.~V. Gambuzza, A. Buscarino, S. Chessari, L. Fortuna, R. Meucci, and M.
  Frasca, Phys. Rev. E {\bf 90},  032905  (2014).

\bibitem{ROS14a}
D.~P. Rosin, D. Rontani, N.~D. Haynes, E. Sch{\"o}ll, and D.~J. Gauthier, Phys.
  Rev.~E {\bf 90},  030902(R)  (2014).

\bibitem{LAR15}
L. Larger, B. Penkovsky, and Y. Maistrenko, Nature Commun. {\bf 6},  7752
  (2015).

\bibitem{MOT13a}
A.~E. Motter, S.~A. Myers, M. Anghel, and T. Nishikawa, Nature Phys. {\bf 9},
  191  (2013).

\bibitem{GON14}
J.~C. Gonzalez-Avella, M.~G. Cosenza, and M.~S. Miguel, Physica~A {\bf 399},
  24  (2014).

\bibitem{RAT00}
N.~C. Rattenborg, C.~J. Amlaner, and S.~L. Lima, Neurosci. Biobehav. Rev. {\bf
  24},  817  (2000).

\bibitem{ROT14}
A. Rothkegel and K. Lehnertz, New J. of Phys. {\bf 16},  055006  (2014).

\bibitem{LOO15}
S. Loos, J.~C. Claussen, E. Sch{\"o}ll, and A. Zakharova,   (2015),
  arXiv:1508.04010v2.

\bibitem{HU93a}
G. Hu, T. Ditzinger, C.~Z. Ning, and H. Haken, Phys.~Rev.~Lett. {\bf 71},  807
  (1993).

\bibitem{PIK97}
A. Pikovsky and J. Kurths, Phys.~Rev.~Lett. {\bf 78},  775  (1997).

\bibitem{NEI97}
A.~B. Neiman, P.~I. Saparin, and L. Stone, Phys.~Rev.~E {\bf 56},  270  (1997).

\bibitem{LIN04}
B. Lindner, J. Garc{\'i}a-Ojalvo, A.~B. Neiman, and L. Schimansky-Geier,
  Phys.~Rep. {\bf 392},  321  (2004).

\bibitem{SIE14c}
J. Sieber, O.~E. Omel'chenko, and M. Wolfrum, Phys. Rev. Lett. {\bf 112},
  054102  (2014).

\bibitem{BIC15}
C. Bick and E.~A. Martens, New J.~Phys. {\bf 17},  033030  (2015).

\bibitem{FIT61}
R. FitzHugh, Biophys. J. {\bf 1},  445  (1961).

\bibitem{NAG62}
J. Nagumo, S. Arimoto, and S. Yoshizawa., Proc. IRE {\bf 50},  2061  (1962).

\bibitem{SCO75}
A. Scott, Rev. Mod. Phys. {\bf 47},  487  (1975).

\bibitem{KLI15}
V. Klinshov, L. L{\"u}cken, D. Shchapin, V.~I. Nekorkin, and S. Yanchuk, Phys.
  Rev. Lett. {\bf 114},  178103  (2015).

\bibitem{CIS03}
M. {Ciszak}, O. Calvo, C. Masoller, C.~R. Mirasso, and R. Toral, Phys. Rev.
  Lett. {\bf 90},  204102  (2003).

\bibitem{ROS11a}
D.~P. Rosin, K.~E. Callan, D.~J. Gauthier, and E. Sch{\"o}ll, Europhys. Lett.
  {\bf 96},  34001  (2011).

\bibitem{SHI04}
S.~I. Shima and Y. Kuramoto, Phys. Rev.~E {\bf 69},  036213  (2004).

\bibitem{HEI10}
M. Heinrich, T. Dahms, V. Flunkert, S.~W. Teitsworth, and E. Sch{\"o}ll,
  New~J.~Phys. {\bf 12},  113030  (2010).

\bibitem{KOZ98}
R. Kozma, Phys. Lett.~A {\bf 244},  85  (1998).

\bibitem{HUL04}
E. Hulata, I. Baruchi, R. Segev, Y. Shapira, and E. Ben-Jacob, Phys.~Rev.~Lett.
  {\bf 92},  198105  (2004).

\bibitem{HEN11}
J.~A. Henderson and P.~A. Robinson, Phys. Rev. Lett. {\bf 107},  018102
  (2011).

\bibitem{DUB99}
J.~L.~A. Dubbeldam, B. Krauskopf, and D. Lenstra, Phys.~Rev.~E {\bf 60},  6580
  (1999).

\bibitem{GIA00}
G. Giacomelli, M. Giudici, S. Balle, and J.~R. Tredicce, Phys.~Rev.~Lett. {\bf
  84},  3298  (2000).

\bibitem{AVI04}
J.~F.~M. Avila, H.~L.~D. de~S.~Cavalcante, and J.~R.~R. Leite, Phys.~Rev.~Lett.
  {\bf 93},  144101  (2004).

\bibitem{ZIE13}
D. Ziemann, R. Aust, B. Lingnau, E. Sch{\"o}ll, and K. L{\"u}dge,
  Europhys.~Lett. {\bf 103},  14002  (2013).

\bibitem{HIZ06}
J. Hizanidis, A.~G. Balanov, A. Amann, and E. Sch{\"o}ll, Phys.~Rev.~Lett. {\bf
  96},  244104  (2006).

\bibitem{HUA14}
Y. Huang, H. Qin, W. Li, S. Lu, J. Dong, H.~T. Grahn, and Y. Zhang, EPL {\bf
  105},  47005  (2014).

\bibitem{JAN03}
N.~B. Janson, A.~G. Balanov, and E. Sch{\"o}ll, Phys.~Rev.~Lett. {\bf 93},
  010601  (2004).

\bibitem{USH05}
O.~V. Ushakov, H.~J. W{\"u}nsche, F. Henneberger, I.~A. Khovanov, L.
  Schimansky-Geier, and M.~A. Zaks, Phys.~Rev.~Lett. {\bf 95},  123903  (2005).

\bibitem{ZAK10a}
A. Zakharova, T. Vadivasova, V. Anishchenko, A. Koseska, and J. Kurths, Phys.
  Rev. E {\bf 81},  011106  (2010).

\bibitem{ZAK13}
A. Zakharova, A. Feoktistov, T. Vadivasova, and E. Sch{\"o}ll, Eur. Phys. J.
  Spec. Top. {\bf 222},  2481  (2013).

\bibitem{GEF14}
P.~M. Geffert, A. Zakharova, A. V{\"u}llings, W. Just, and E. Sch{\"o}ll, Eur.
  Phys. J.~B {\bf 87},  291  (2014).

\bibitem{SEM15}
V. Semenov, A. Feoktistov, T. Vadivasova, E. Sch{\"o}ll, and A. Zakharova,
  Chaos {\bf 25},  033111  (2015).

\bibitem{ROS09b}
O.~A. Rosso and C. Masoller, Phys. Rev. E {\bf 79},  040106(R)  (2009).


\bibitem{WOL11a}
M. Wolfrum, O.~E. Omel'chenko, S. Yanchuk, and Y. Maistrenko, Chaos {\bf 21},
  013112  (2011).

\bibitem{HAU15}
S.~W. Haugland, L. Schmidt, and K. Krischer, Sci. Rep. {\bf 5},  9883  (2015).

\bibitem{ZAK14}
A. Zakharova, M. Kapeller, and E. Sch{\"o}ll, Phys.~Rev.~Lett. {\bf 112},
  154101  (2014).

\end{thebibliography}

\end{document}